# Optical orientation of the homogeneous non-equilibrium Bose-Einstein condensate of bright excitons (polaritons)

V.L. Korenev

*A. F. Ioffe Physical Technical Institute, St. Petersburg, 194021 Russia*

*A simple model, describing the steady state of the non-equilibrium polarization of a homogeneous Bose-Einstein condensate of exciton polaritons, has been considered. It explains the suppression of spin splitting of a non-equilibrium polariton condensate in an external magnetic field, the linear polarization, the linear-to-circular polarization conversion and the unexpected sign of circular polarization of the condensate on equal footing. It has been shown that inverse effects are possible, to wit, the spontaneous circular polarization and the enhancement of spin splitting of a non-equilibrium condensate of polaritons.*

Artificially created systems of bosons are attractive for studying the different aspects of Bose-Einstein condensation [1]. The condensates of indirect excitons in quantum wells [2] and bright excitons (polaritons) in microcavities [3, 4] are unique, since their properties can be controlled optically. In spite of the non-equilibrium character of optical creation a quasi-equilibrium polariton condensate still can be formed close to the momentum k=0 [4]. The states of excitons and polaritons are characterized by projections of the angular momentum ±1 on the growth axis of the well and can be described in terms of a fictive spin - pseudospin ½ [5]. Recently [6] the suppression of spin splitting of a polariton condensate has been observed in a continuous wave (CW) regime. The sign of circular polarization of condensate appeared to be opposite to that of equilibrium condensate considered in previously in [7]. In turn, short lifetime of polaritons (a few ps) points to the strong deviation of the spin system from equilibrium. Only spin dynamics of non-equilibrium condensate was considered up to now [8]. However a self-consistent explanation of the CW experiment [6] is absent.

Here a simple model of non-equilibrium spin polarization of homogeneous ensemble of polaritons is considered in CW regime. This model makes it possible to describe all the principal experimental results, such as the suppression of Zeeman splitting of a non-equilibrium condensate and the unexpected sign of circular polarization of polaritons within a solely non-equilibrium approach. It predicts also the inverse effects: the spontaneous circular polarization and the enhancement of spin splitting of a non-equilibrium condensate of polaritons in a zero external magnetic field.

A model of the pseudospin ½ for an ensemble of isolated bright excitons was worked out in Refs. [5]. In the absence of spin relaxation the states of bright and dark excitons in a GaAs-type quantum well can be divided into two uncorrelated two-level systems. Each of the systems can be considered as a quasi-particle with pseudospin ½. For the case of bright excitons it is natural to choose the states with momentum projection $|+1\rangle$ and $|-1\rangle$ on the growth axis of the well, as a projection of the pseudospin "up" $|z,+1/2\rangle$ and "down" $|z,-1/2\rangle$, respectively, on the z-axis in the effective three-dimensional space of the pseudospin $(x, y, z)$ (one shouldn't confuse the axes of pseudospin space with the axes in the real space). In a state $|+1\rangle$ ($|-1\rangle$) the exciton emits the circularly-polarized light in $\sigma^+$ ($\sigma^-$) polarization. Then the z-component $S_z$ of a mean (over ensemble of excitons) pseudospin determines the degree of circular polarization of luminescence $P_c = 2S_z$. The state $|X\rangle = (|+1\rangle + |-1\rangle)/\sqrt{2}$ is dipole-active along the crystallographic axis $[1\bar{1}0]$. It is described by x-component of the pseudospin $|x, +1/2\rangle$. The state $|Y\rangle = (|+1\rangle - |-1\rangle)/\sqrt{2}i$, dipole-active along the orthogonal axis [110], is described by component $|x, -1/2\rangle$. Then the *x*-projection of the mean pseudospin determines the degree of linear polarization of exciton luminescence relative to crystal axes $[1\bar{1}0]/[110]$: $P_l = 2S_x$. Finally, the states $|X'\rangle = (|X\rangle + |Y\rangle)/\sqrt{2}$ and $|Y'\rangle = (-|X\rangle + |Y\rangle)/\sqrt{2}$ are polarized along directions [100] и [010], respectively. They correspond to two possible projections of pseudospin on the *y*-axis: $|y, +1/2\rangle$ and $|y, -1/2\rangle$. Consequently, the *y*-component of the mean pseudospin determines the degree of linear polarization of exciton luminescence relative to crystallographic axes $[100]/[010]$: $P_{l'} = 2S_y$. All three components of the mean pseudospin can be measured, since they are connected unambiguously with three Stokes parameters characterizing the luminescence polarization:

$$P_c = 2S_z; \qquad P_l = 2S_x; \qquad P_{l'} = 2S_y \qquad (1)$$

Relations (1) between light polarization and pseudospin admit a simple geometric interpretation. As is known, the light polarization is characterized by three Stokes parameters, which can also be considered as a vector with coordinates $\vec{P} = (P_l, P_{l'}, P_c)$ in the three-dimensional space. The end of this vector runs through the whole multitude of polarization values and at a given value of polarization describes a sphere, which is called the Poincare sphere. The pseudospin of an ensemble of excitons can be conceived as a vector $\vec{S}$ in the three-dimensional space. By virtue of (1), each position of pseudospin in the effective three-dimensional space corresponds here to a definite position of vector $\vec{P}$ on the Poincare sphere. Given orientation of the pseudospin of excitons fully determines the polarization properties of exciton luminescence and vice versa.

In order to describe the time evolution of exciton polarization and luminescence we must know the Hamiltonian of excitons. In the case of symmetry $C_{2v}$ (an asymmetric well, grown in the direction [001]) the main axes of anisotropic electron-hole exchange interaction in an exciton coincide with crystal axes [110] and [1$\bar{1}$0]. The Hamiltonian of a bright exciton in magnetic field in Faraday geometry $\vec{B}\|[001]$ can be written in terms of pseudospin as

$$\hat{H} = \frac{\hbar}{2}(\omega_b \hat{\sigma}_x + \Omega_{ext}\hat{\sigma}_z) = \frac{\hbar}{2}\vec{\Omega}\cdot\vec{\sigma}, \qquad (2)$$

where $\sigma_i$ are Pauli matrices in $x, y, z$ axes of the effective space that the pseudospin rotates in, $\Omega_{ext} = \mu_B g_\| B/\hbar$ is the Larmor precession frequency of the pseudospin in an external field ( $g_\|$ is the longitudinal g-factor of the exciton). In the absence of magnetic field ($\Omega_{ext} = 0$) the components of the radiative doublet are polarized along the axes $[1\bar{1}0]$ and [110] with the energy of splitting $\delta_b = \hbar\omega_b$. The pseudospin dynamics is described by the Bloch equation [9,10]

$$\frac{d\vec{S}}{dt} = \frac{\vec{S}^0 - \vec{S}}{\tau_b} + \vec{\Omega}\times\vec{S} - \frac{\vec{S}-\vec{S}_T}{\tau_s} \qquad (3)$$

where on the right hand side the first term describes the generation and recombination of the pseudospin, the second term, Larmor precession with frequency $\vec{\Omega} = (\omega_b, 0, \Omega_{ext})$. The spin relaxation is included phenomenologically [10] by adding the Bloch equation with term (the third term) describing the evolution of a pseudospin with characteristic time $\tau_s$ towards an equilibrium value $\vec{S}_T = -\frac{1}{2}\frac{\vec{\Omega}}{\Omega}th\left(\frac{\hbar\Omega}{2T}\right)$ at a temperature $T$. Such a description is reasonable in the case when there occurs a flip of the exciton spin as a whole (that is, from the $|+1\rangle$ state into the $|-1\rangle$ state and vice versa), so that the system remains a two-level one, and the concept of quasiparticles with half-integer pseudospin is still valid. According to Eq. (2), the pseudospin vector $\vec{S}$ rotates about vector $\vec{\Omega} = (\omega_b, 0, \Omega_{ext})$ with Larmor frequency $\Omega = \sqrt{\omega_b^2 + \Omega_{ext}^2}$. If the time $\tau_b, \tau_S$ of exciton is sufficiently long, so that $\Omega\tau_b, \Omega\tau_S \gg 1$, then the Larmor precession about the $\vec{\Omega}$ vector depolarizes the components of the initial pseudospin $\vec{S}^0 = (S_x^0, S_y^0, S_z^0)$ transversal to $\vec{\Omega}$, while the component along $\vec{\Omega}$ is conserved. As a result, the stationary orientation of pseudospin is attained by projection of $\vec{S}^0$ onto the direction of $\vec{\Omega}$.

Whilst only isolated excitons were put under consideration, there arose no question of inter-particle coupling. At the same time, as has been shown in Ref. [11], even in a diluted system of excitons with concentration $N_x \ll 1/\pi a_x^2$ (within the circle of Bohr radius of the exciton $a_x$ there are no other excitons [12]) there occurs their spin-spin interaction with each other. It has been found that under creation of excitons in GaAs/AlGaAs quantum well by the pulse of circularly-polarized light, the luminescence spectrum consists of two lines, corresponding to recombination of excitons from states $|+1\rangle$ and $|-1\rangle$. The energy splitting is proportional to the number of excitons as well as to the degree of their circular polarization. This splitting has been explained as due to the exchange interaction between electrons, which lifts degeneracy of radiative doublet. The spin-spin interaction of radiative excitons within the framework of the model of quasiparticles with pseudospin ½ is described by anisotropic exchange interaction [13, Appendix]. In the mean field approximation (MFA) the spin splitting of an exciton results from the action of an effective exchange magnetic field along $z$, and proportional to mean pseudospin $S_z = \frac{1}{2}\langle \sigma_j \rangle$ of an exciton ensemble. The MFA spin Hamiltonian of a given exciton takes the form

$$\hat{H}_{MFA} = -\frac{J}{2}\hat{\sigma}_z S_z$$

The evolution of exciton polarization under the action of Hamiltonian $H_{MFA}$ can be considered as a precession $\vec{S}$ with angular frequency $\vec{\Omega}'(\vec{S}) = (0, 0, -\omega_{exch} S_z)$, where $\omega_{exch} = J/\hbar$. Generalization of precession dynamics for the case of interaction between excitons consists in the addition of the term $\vec{\Omega}'(\vec{S}) \times \vec{S}$. Then the total Larmor frequency of the pseudospin

$$\vec{\Omega} = (\omega_b, 0, \Omega_{ext} - \omega_{exch} S_z) \qquad (4)$$

depends on the orientation of exciton ensemble.

The diluted gas of bright excitons forms a set of quasiparticles with pseudospin ½, which obey Bose statistics (the projection of their angular momentum ±1 is integer). The Bose statistics of excitons will manifest itself at fairly low temperatures, when de Broglie wave length $\lambda_{Br}$ of excitons becomes comparable with the characteristic distance between them $N_x \pi \lambda_{Br}^2 \geq 1$ (under fulfillment of the condition of dilution, $N_x \pi a_x^2 << 1$). This situation takes place in the system of indirect excitons [2] and in the semiconductor microcavities [3] with quantum wells. In the last case, under conditions of strong coupling of photons and excitons there arises a hybrid mode, a polariton with momentum projection +1 and −1 on the growth axis. Of course, we can also treat the polariton as a quasiparticle with pseudospin ½ and apply the above considerations to the ensemble of polaritons. The pseudospin concept was successfully applied to cavity polaritons in [14] using the exchange term $H_{MFA}$ to describe the effect of self-induced Larmor precession of polariton pseudospin. At a temperature below the critical $T_c$ the Bose condensation of polaritons takes place [3], and in such a case the order parameter will be not a scalar, but a vector function, defining the polarization of condensate [4]. The complex order parameter can be expressed through the density components of the pseudospin $\vec{S}$ of the condensate. In spite of the non-equilibrium character of optical creation of ensemble, a quasi-equilibrium polariton condensate can be formed close to the momentum k=0 [4]. However, due to very short lifetime of polaritons (a few ps) the spin degree freedom is essentially non-equilibrium. It was shown [8] that without spin relaxation term the pseudospin dynamics $\vec{S}$ of non-equilibrium polariton condensate is still determined by the precession equation (3) in which the vector $\vec{\Omega}(\vec{S})$ is given by Eq.(4), with the quantities in Eq.(4) having similar meaning as for the case of excitons. Ref.[8] solved the dynamic Equations (3,4) for $\tau_S = \infty$. Nontrivial is the question of spin relaxation of polaritons in the condensate. The fact is that any finite splitting of spin levels at a temperature below $T_c$ brings about an increased equilibrium population of the lower state as compared with Boltzmann distribution due to Bose collectivization [13, Appendix]. Therefore it can be assumed with good accuracy that the equilibrium condensate of polaritons is characterized by the maximum possible polarization. Thus, within the simplest model the pseudospin dynamics of a homogeneous nonequilibrium-in-spin condensate of polaritons at $T < T_c$ is described by Bloch equation (3), in which the equilibrium value

$$\vec{S}_T = -\frac{1}{2}\frac{\vec{\Omega}}{\Omega} \qquad (5)$$

despite the fact that $T \neq 0$ (but $T < T_c$).

Here I solve the Equations (3-5) in the steady state regime in the limit $\Omega\tau_b, \Omega\tau_s >> 1$ when the non-equilibrium pseudospin turns many times about the vector $\vec{\Omega}$, and only its projection onto direction $\vec{\Omega}$ remains in the stationary state:

$$\vec{S} = \frac{\tau_s}{\tau+\tau_s}\frac{(\vec{\Omega}\vec{S}^0)\vec{\Omega}}{\Omega^2} - \frac{\tau}{\tau+\tau_s}\frac{\vec{\Omega}}{2\Omega} \qquad (6)$$

Zeeman splitting of pseudospin sublevels

$$E_Z = \sqrt{\omega_b^2 + (\Omega_{ext} - \omega_{exch}S_z)^2} \qquad (7)$$

Let the initial pseudospin of an ensemble of polaritons $\vec{S}^0$ be directed along the *x*-axis (Fig. 1A). It is seen from the Fig.1 that vector $\vec{\Omega}$ lies in the plane $(x,z)$ and forms an angle $\arctan([\Omega_{ext} - \omega_{exch}S_z]/\omega_b)$ with the *x*-axis. The dependencies $P_c(\Omega_{ext})$, $P_\ell(\Omega_{ext})$ and $\Delta E_Z(\Omega_{ext})$ in Fig. 1 are calculated from the Eqs. (1, 4, 6, 7) with parameters $\omega_{exch} = 8\omega_b$, $\tau_s = 2.5\tau_b$. It can be seen that a longitudinal magnetic field $\Omega_{ext} \neq 0$ brings about the following effects: (i) the exchange interaction of polaritons compensates Zeeman splitting in an external magnetic field up to $\Omega_{ext} < 2\omega_b$ (Fig.1B); (ii) the linear polarization of polaritons $S_x$ decreases (Fig. 1C); (iii) there appears a circular polarization of polaritons, i.e. $S_z \neq 0$ (Fig.1D); (iv) in a weak field $\Omega_{ext} < 2\omega_b$ (Fig.1D) the sign of the nonequilibrium part of circular polarization of polaritons $S_z$ coincides with the sign of $\Omega_{ext}$, i.e. is opposite to the sign of the equilibrium polarization of polaritons $S_T$. All these features were observed experimentally [6].

Earlier the items (i-iii) were predicted for the spin-equilibrium condensate. Ref. [7] calculated the zero-temperature equilibrium spin $\vec{S}_T = -\frac{1}{2}\frac{\vec{\Omega}}{\Omega}$ where $\vec{\Omega}(\vec{S}_T)$ is given by Eq.(4). According to [7], however, the sign of equilibrium circular polarization $S_{Tz}$ (item iv) must be opposite to the sign of $\Omega_{ext}$ for *any* $\Omega_{ext}$ in clear contradiction with experiment [6]. Moreover short lifetime (a few ps) of cavity polaritons points to the fact that the spin system should be considered as strongly non-equilibrium from the very outset. Therefore the equilibrium approach [7] cannot provide a self-consistent description of experiment.

Note that for explanation of the experimental results [6] on polarization and quenching of Zeeman splitting, the exchange constant should be ferromagnetic, $J>0$. Indeed, the Larmor precession frequency in the exchange field is determined by the product $J \cdot S_z$ that should be positive for the compensation (Fig.1 a,b). To determine the $J$ sign the sign of $S_z$ should be known. The same experiment [6] suggests that the $S_z$ sign in the region of compensation is opposite to the equilibrium sign of $S_{Tz}$, which is negative. Hence $S_z > 0$ and one obtains immediately $J>0$ [15]. Constant $J$ originates mainly from the exchange interaction of the carriers of the same name but belonging to different excitons and from the exciton-exciton exchange [16, 17]. Also the interaction of polaritons with dark excitons and electron-hole plasma (under non-resonant excitation) may contribute to the constant $J$. Generally there are no principal limitations as to the sign of $J$.

It should be noted that the suppression of spin splitting is not complete ($E_Z \geq \hbar\omega_b$) in apparent contradiction with experiment [6]. The fact is that the measurements of splitting were carried out in crossed circular polarizers. In a weak magnetic field, however, two orthogonal spin states are polarized elliptically. Strictly speaking, the PL polarization analyzers should be also elliptical, but this would be inconvenient from the practical point of view. Nevertheless, the question of non-zero splitting is principal, since an explanation has recently appeared [18], predicting the complete quenching of splitting, i.e. $E_z = 0$. The given model can be checked easily in a zero field, in which two states of the condensate are linear-polarized orthogonally. Therefore the PL spectra of the condensate must show a splitting of $E_Z = \hbar \omega_b$ in two crossed linear polarizations.

The initial pseudospin $\vec{S}^0$ is determined by two reasons: (i) polarization of incident light [5], with $\vec{S}^0$ components being related to Stokes parameters $P_l^0, P_{l'}^0, P_c^0$ of the exciting light similar to Eq.(1); (ii) the anisotropy of the optical properties of nanostructure leading to the magnetic field independent linear dichroism effect and the anisotropy of recombination [9]. Interestingly, in the latter case the "initial" pseudospin $\vec{S}^0$ can be accumulated in the process of thermalization due to the polarization-selective escape of the cavity. In case (i) the $\vec{S}^0$ can be manipulated through the laser polarization [5, 9, 19, 20, 21]. In contrast to it, in case (ii) the vector $\vec{S}^0$ is fixed by the specific structure geometry. Nevertheless, one can change the orientation of $\vec{S}^0$ via externally applied uniaxial stress. The reversal of the $\vec{S}^0$ direction converts the suppression of Zeeman splitting into its enhancement. Even the spontaneous circular polarization of polaritons may take place in a non-equilibrium condensate. Indeed, let the initial pseudospin of polaritons $\vec{S}^0$ be directed opposite to the x-axis. As is seen from Fig.2A, in this case the projection $S_z$ onto $\Omega_{ext}$ is negative, and the effective exchange field will be added to the external one. It will result in enhancement of Zeeman splitting rather than suppression. It can be said that negative feedback converts to positive. This leads to the multivalued dependence $\vec{S}(\Omega_{ext})$ in Fig.2, with the stability of stationary states being determined with the help of dynamic Eq. (3). Stable (unstable) states are shown by solid (dashed) lines in Fig.2. If the exchange constant is sufficiently large, a spontaneous circular polarization $S_z \neq 0$ will appear (Fig.2D), sustained by the exchange field. Such a state will be characterized by an appreciable Zeeman splitting in the exchange field even at $\Omega_{ext} = 0$ (Fig.2B) [22].

Thus, a simple model has been analyzed, which describes the steady state regime of the spin non-equilibrium homogeneous Bose-Einstein condensate of exciton polaritons. It explains all basic experimental results on the condensate polarization as well as the suppression of Zeeman splitting together with the unusual sign of circular polarization. In turn, it predicts inverse effects, to wit, the spontaneous circular polarization in a zero external magnetic field and the enhancement of Zeeman splitting of a non-equilibrium condensate.

The author is deeply grateful to M.M. Glazov, A.V. Kavokin, K.V. Kavokin for valuable discussions. The work has been partially supported by RFBR and programs of Russian Academy of Sciences.

*APPENDIX:*

Selected points from the Chapter IV of Habilitation thesis

# Magnetic interactions in spin systems of semiconductor heterostructures

V.L. Korenev

Condensed matter physics, St Petersburg

Electronic version has been completed 7 April 2003

## Chapter IV. Fine structure of zero-dimensional excitons (page 101)

*Pages 116-117.*

Above we have considered in detail the two-level system of bright excitons as quasiparticles with pseudospin ½. All the results obtained could be deduced from the Bloch equation for the exciton pseudospin:

$$\frac{d\vec{S}}{dt} = \frac{\vec{S}^0 - \vec{S}}{\tau_b} + \vec{\Omega} \times \vec{S} \qquad \text{(IV.10)}$$

Components of the frequency $\vec{\Omega}$ follow from the Hamiltonian $\hat{H} = \frac{\hbar}{2}\vec{\Omega}\cdot\hat{\boldsymbol{\sigma}} = \frac{\hbar}{2}\left(\omega_b\hat{\sigma}_x + \tilde{\omega}_b\hat{\sigma}_y + \Omega_{ext}\hat{\sigma}_z\right)$, which determines the fine structure of the bright excitons: $\vec{\Omega} = \left(\omega_b, \tilde{\omega}_b, \Omega_{ext}\right)$. Steady state solution of the Eq. (IV.10) taking into account (IV, 1)

$$P_c^0 = 2S_z^0; \qquad P_l^0 = 2S_x^0; \qquad P_{l'}^0 = 2S_y^0 \qquad \text{(IV, 11)}$$

$$P_c = 2S_z; \qquad P_l = 2S_x; \qquad P_{l'} = 2S_y$$

relates the photoluminescence polarization $\vec{P}$ with polarization of the exciting light $\vec{P}^0$.

# Pages 120-126
## *IV.2.* Collective effects in an ensemble of excitons

We have introduced in section IV.1 a model that allows us to consider the exciton as a quasiparticle with the pseudospin ½. As long as we considered a single exciton, the question of statistics of such quasiparticles did not arise. However, in experiments with an exciton ensemble of sufficiently high density, statistics may prove to be important. The term itself "the quasiparticle with the half-integral pseudospin" can be misleading, since a wish then arises to ascribe them the Fermi statistics, which is incorrect. The fact is that an exciton is formed by two fermions, an electron and a hole. The formation from two fermions will have Bose-statistics. In this the difference between the pseudospin and the real spin manifests itself spectacularly. The term "pseudospin ½" follows from the fact that the properties of a two-level system are in many respects analogous to the properties of a single electron with the half-integral spin. Bright excitons create a two-level system of quasi-particles with the fictitious spin ½, obeying however Bose-statistics (with integer angular momentum projection $\pm 1$).

All said above makes sense, if a two-dimensional exciton retains its individuality (the electron-hole bound state) with a good accuracy in the presence of other quasiparticles. To this end the fulfillment of the condition $N_x \pi a_x^2 \ll 1$ is required, i.e., in the circle of the Bohr radius of an exciton there must not be any other excitons. In that case the ensemble of excitons may be thought of, to a good degree of accuracy, as the gas of quasiparticles interacting weakly with each other. The ground state of an individual 1s exciton constitutes a quartet of levels characterized by the projection of the full moment M (electron and hole) onto the growth axis of the quantum well. The axially symmetrical exchange interaction of an electron and hole splits that state into a radiative doublet (M = ± 1) and a pair of non-radiative states with M = ± 2. If the non-radiating states are neglected (e.g. spin relaxation is suppressed) then the system becomes a two-level one, and the exciton can be considered as a quasi-particle with the pseudo-spin ½. Bose statistics will show up at sufficiently low temperatures and high concentrations, when the de Broglie wavelength of excitons becomes comparable with the characteristic distance between them (but still the condition of rarefaction $N_x \pi a_x^2 << 1$ is being fulfilled). The possible manifestation of exciton statistics in experiments was paid attention to in Ref. [2].

On the other hand, as was shown in Ref. [11], even in a rarefied system of excitons their spin-spin interaction takes place. It has been found that under excitation of excitons in a GaAs/AlGaAs quantum well by the pulse of circularly polarized light, the luminescence spectrum consists of two lines, corresponding to recombination of electrons from the states $|+1\rangle$ and $|-1\rangle$. It is remarkable that there is energy-level splitting, which is proportional to both the number of excitons and the degree of their circular polarization. That splitting has been explained by exchange interaction of excitons. Within the pseudospin model framework, the spin splitting of excitons is the result of the pseudospin being acted upon by an effective magnetic field, whose magnitude and direction are determined by the mean pseudospin of the ensemble of excitons. In the given case the ensemble is oriented optically, so the effective magnetic field is directed along the z-axis in the pseudospin space and is proportional to the z-component of the mean pseudospin of excitons. The magnitude of splitting in this field $\Delta E = JS_z$, where $J$ is the constant, characterizing the interaction of bright excitons with each other. The magnitude and sign of this constant is determined by a number of factors, such as exchange interaction of the carriers of the same name but belonging to different excitons, or the change in the inter-exciton electron-hole attraction due to Pauli principle [23]. Besides, there is a long-range exchange interaction of the electron and the hole belonging to different excitons (electrostatic dipole-dipole interaction, which can dominate at low concentrations). Thus the problem of the interacting exciton gas comes to the effective magnetic interaction of Bose particles with pseudospin ½. The shift direction of the spin levels of excitons, according to results in Ref. [11], is such that the constant $J$ is antiferromagnetic ($J < 0$). As to the sign of $J$, however, there are no essential limitations. There exist theoretical prerequisites [16], that under certain conditions the state of electrons with parallel orientation of angular moments (i.e. its sign may be positive) will turn out to be advantageous from the viewpoint of energy. Then the ferromagnetic transition in the system of excitons is possible at a temperature below the critical one. In this case the effect of spontaneous magnetization will be strengthened, in contrast to Fermi systems, owing to tendency of excitons for accumulating in the only one quantum state (Bose statistics). Let illustrate this assertion on the example of a simple model of free excitons in a quantum well with quadratic law of dispersion. In this case the energy spectrum of an exciton consists of two subbands, corresponding to two projections of the pseudospin $s = \pm 1/2$ onto the z-axis:

$$E_{K,s} = \frac{\hbar^2 K^2}{2M} - JS_z s \qquad (IV, 14)$$

where $M$ is the translation mass of the exciton, $K$ is its pulse. The number of quasiparticles $n_{K,s}$ in the $|K,s\rangle$ state is determined by Bose-Einstein statistics:

$$n_{K,s} = \left[\exp\left(\frac{E_{K,s} - \mu}{T}\right) - 1\right]^{-1} \qquad (IV, 15)$$

where temperature $T$ is measured in energy units, $\mu$ is chemical potential, which is given by total concentration of excitons $N$ with projections of pseudospins $|+1/2\rangle$ ($N_+$) and $|-1/2\rangle$ ($N_-$):

$$N = N_+ + N_- = \sum_K \left(n_{K,+1/2} + n_{K,-1/2}\right) \qquad (IV, 16)$$

The average-in-ensemble z-component of the pseudospin of excitons,

$$S_z = \frac{1}{2}\frac{N_+ - N_-}{N_+ + N_-} \qquad (IV, 17)$$

is determined as the result of a self-consistent solution of the system of simultaneous equations (IV, 14-17). It has a non-trivial solution complying with ferromagnetic ordering of the system of pseudospins, if the temperature is below the critical point, $T < T_c$. Let's calculate that temperature, assuming polarization of excitons to be small, i.e. $S_z \ll 1/2$. In that case the splitting of pseudospin subbands is also small, and expression (IV, 17) can be expanding into power series $\Delta E = JS_z$. Restricting ourselves to a linear term, we'll obtain:

$$S_z = \frac{1}{2}\frac{N_+ - N_-}{N_+ + N_-} = -\frac{JS_z}{2}\frac{1}{N}\frac{\partial N}{\partial \mu} + O(S_z^3) \qquad (IV, 18)$$

The non-trivial solution $(S_z \neq 0)$ of equation (IV, 18) takes place under the condition

$$-\frac{J}{2}\frac{1}{N}\frac{\partial N}{\partial \mu} \geq 1, \qquad (IV, 19)$$

coinciding with the well-known Stoner criterion for the zonal ferromagnetism of Fermi systems [24]. In the given case, however, it is necessary, when calculating (IV, 19), to use Bose-Einstein statistics. The relationship between the number of particles and the chemical potential can be found by elementary integration:

$$N = \sum_{K,s} n_{K,s} = 2\int \frac{d^2\vec{K}}{(2\pi)^2}\frac{1}{\exp\left(\frac{E_K - \mu}{T}\right) - 1} = -\frac{MT}{\pi\hbar^2}\ln\left(1 - e^{\frac{\mu}{T}}\right) \qquad (IV, 20)$$

Substituting (IV, 20) into (IV, 19), we obtain the condition for ferromagnetism of the exciton gas, which determines the temperature of transition $T_c$:

$$\frac{JM}{2\pi\hbar^2 N}\left[\exp\left(\frac{\pi\hbar^2 N}{MT_c}\right) - 1\right] = 1 \qquad (IV, 21)$$

Here an important role belongs to the parameters $\pi\hbar^2 N/MT_c = \pi N \lambda_{br}^2$, which characterizes the number of excitons in a circle with radius equal to de Broglie wavelength: if the radius is more than unity in the transition point, then the transition will depend comparatively weakly on the magnitude of the constant of exchange coupling between pseudospins. Indeed, neglecting the unity in square brackets of Eq. (IV, 21) we obtain

$$T_c = \frac{\pi\hbar^2 N}{M}\left[\ln\left(\frac{2\pi\hbar^2 N}{JM}\right)\right]^{-1} \qquad (IV, 21)$$

As follows from Eq. (IV, 21), the transition temperature increases with growth of $J$ but only logarithmically. That's why the phase transition can occur even when the constant of exchange coupling is rather weak.

So we have restricted the problem of interaction in a rarified exciton gas to a one-particle problem of interaction of a quasiparticle with pseudospin ½ and an effective magnetic field (mean field approximation) proportional to the z-component of the mean pseudospin of excitons. In that case the one-particle spin Hamiltonian takes the form

$$\hat{H} = -\frac{J}{2}\hat{\sigma}_z S_z \qquad (IV, 22)$$

As has been mentioned above, an ensemble of bright excitons is characterized by three components of the mean pseudospin. Therefore under arbitrary polarization of the exciton ensemble the one-particle Hamiltonian can be represented in the form of anisotropic exchange interaction of a single exciton with an effective magnetic field whose magnitude and direction are determined by the magnitude and direction of the (x, y, z)-components of the mean-in-ensemble pseudospin:

$$\hat{H} = -\frac{J}{2}\hat{\sigma}_z S_z - \frac{J_\perp}{2}(\hat{\sigma}_x S_x + \hat{\sigma}_y S_y) \qquad (IV, 23)$$

The phenomenological constant $J_\perp \neq J$ is introduced here from symmetry considerations. The anisotropy of the Hamiltonian (IV, 23) follows from the principal distinction between the (x,y)-components of pseudospin and the z-component: the transverse components describe spin correlation of an electron and hole in an exciton (linearly polarized dipoles) and don't change the sign on time inversal, while the z-component characterizes the difference in populations of exciton levels plus- and minus unity (circularly polarized dipoles). It changes the sign on time inversal. The evolution of exciton polarization under the influence of the Hamiltonian (IV, 23) can be represented clearly as pseudospin precession with angular frequency $\vec{\Omega}'(\vec{S}) = (J_\perp S_x, J_\perp S_y, J S_z)/\hbar$. Generalization of Bloch equation (IV, 10) to the case of interaction between excitons reduces to the addition of a summand $\vec{\Omega}'(\vec{S}) \times \vec{S}$, and the system becomes significantly nonlinear. As it seems, the precession of the polarization vector with frequency $\vec{\Omega}'(\vec{S})$ depending on light polarization, has been observed quite recently [25] in experiments on stimulated scattering of excitons (polaritons) in microresonators. However, to identify the given effect unambiguously, further experiments are required.

So, in this section we have formulated a model, according to which light excitons can be considered as quasiparticles with pseudospin ½, obeying Bose-Einstein statistics. This model gives a clear description of the results of a great number of polarization experiments It makes it possible to take into account quite simply the effects of spin relaxation of excitons, dichroism (linear and circular) and the exciton-exciton interaction of light excitons in rarified gas.

I am grateful to Mrs N.Vassilyeva for the translation.

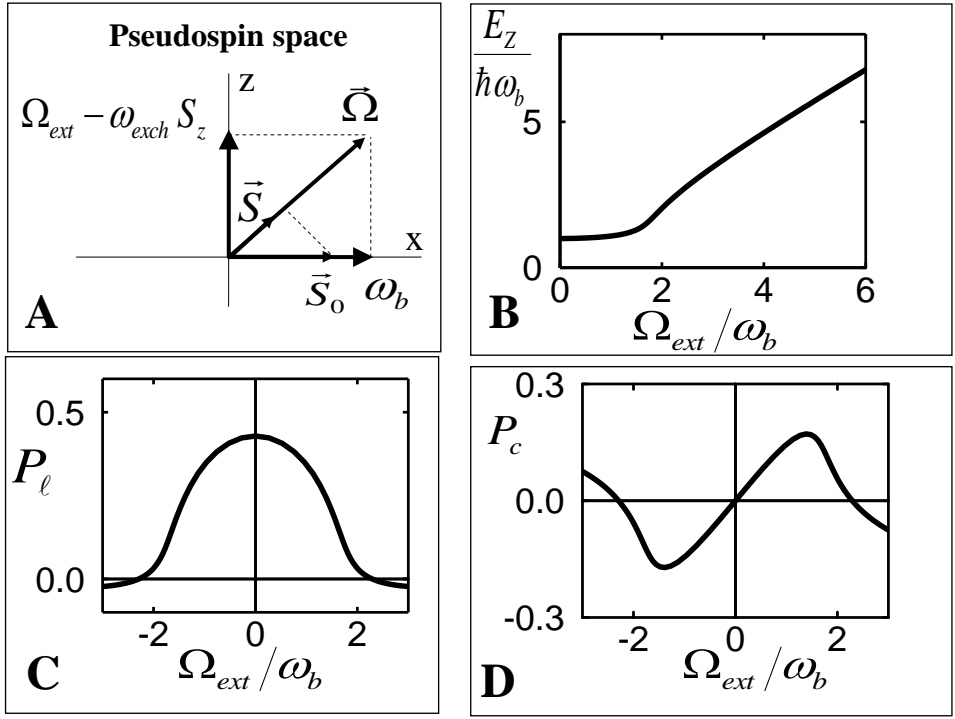

Figure 1. (A) Orientation of the pseudospin vector under stationary conditions in an external magnetic field $(\Omega_{ext} \neq 0)$ and in effective magnetic fields of anisotropic electron-hole coupling in an exciton $(\omega_b \neq 0)$ and of inter-particle spin-spin exchange $(\omega_{exch} \neq 0)$. (B). The Zeeman splitting of the levels $\pm 1/2$ (i.e. the states with angular momentum projection $\pm 1$) vs. $\Omega_{ext}$ frequency. The dependences of the degrees of linear (C) and circular (D) polarization of polariton luminescence on $\Omega_{ext}$ frequency.

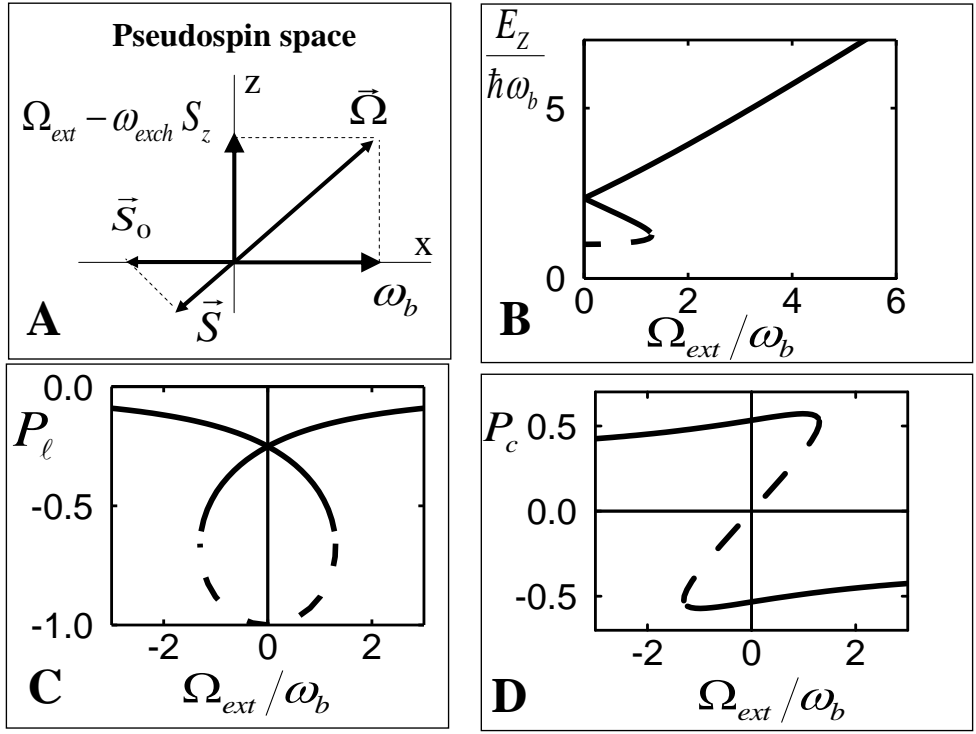

Figure 2. (A) Orientation of the pseudospin vector under stationary conditions in when initial spin $\vec{S}^0$ is opposite to $x$-axis (excitation with [110]-polarized light). (B). The enhanced Zeeman splitting of the levels $\pm 1/2$ vs. $\Omega_{ext}$ frequency. The dependences of the degrees of linear (C) and circular (D) polarization of polariton luminescence on $\Omega_{ext}$ frequency. Solid (dashed) lines show stable (unstable) states.

affect the precessional term in Eq.(3) we conclude that the effective exchange constant $J$ for the Larmor frequency is $J = J_{\parallel} - J_{\perp}$. The anisotropy of the Hamiltonian follows from the principal distinction of the (*x,y*) components of pseudospin from the *z*-component: the transversal components describe correlation of the electron- and hole spin in a polariton (linearly polarized dipoles) and do not change their sign under time inversion. The *z*-component characterizes the difference in population of polariton levels ±1 and changes its sign under time reversal.